# Effects of Diode Laser Photobiomodulation on Peri-Implant Inflammation and Stability in Orthodontic Mini-Implants: A Randomized Controlled Trial

Jun Liu, MDS,[1,2] Linlin Li, MDS,[3] Xiaofei Sun, BA,[4] Qiang Zhang, DDS,[1,2]

[1] *Department of Orthodontics, the Affiliated Hospital of Qingdao University, Qingdao, China.*
[2] *School of Stomatology of Qingdao University, Qingdao, China.*
[3] *Department of Orthodontics, Keen Stomatological Hospital, Weifang, Shandong, China.*
[4] *Vancouver School of Economics, University of British Columbia, Vancouver, Canada.*

## Abstract

Peri-implant inflammation in orthodontic mini-implant may lead to patient discomfort and treatment failure. This study aims to evaluate the effects of diode laser application on the health of mini-implant, preventing peri-implantitis and promoting healing. A randomized controlled trial was conducted involving 30 orthodontic patients (12 males and 18 females, aged 18-32) who had mini-implants implanted on both sides of the maxilla for anterior teeth retraction. One side of each patient was assigned to either an experimental group receiving diode laser irradiation (650 nm, 25 mW) at specific postoperative intervals or a control group receiving simulated irradiation. Clinical assessments included plaque index, modified sulcus bleeding index, probing depth, and incidence of peri-implant mucositis and implant mobility, measured at 1, 4, and 12 weeks post-implantation. Additionally, interleukin-1 beta (IL-1β) levels in peri-implant fluid were analyzed via enzyme-linked immunosorbent assay (ELISA). Results indicated that the experimental group exhibited significantly lower plaque indices, sulcus bleeding indices, and probing depths ($p < 0.05$) compared to the control group. Moreover, the experimental group had fewer cases of peri-implant mucositis ($p < 0.05$), while differences in implant stability were not statistically significant ($p > 0.05$). IL-1β levels were consistently lower in the experimental group throughout the study duration ($p < 0.05$). In conclusion, adjunctive diode laser therapy appears to enhance peri-implant health and reduce complications associated with orthodontic mini-implants, suggesting a promising direction for improving patient outcomes in orthodontics. Future research should explore long-term effects and the mechanisms underlying these benefits.

**Key words:** photobiomodulation; low level laser therapy; light-emitting diode

therapy; mini-implant; peri-implantitis

## Introduction

Orthodontic mini-implants have emerged as an important component in contemporary orthodontic treatment, providing stable anchorage for various dental procedures. However, superior therapeutic outcomes depend on the stability of mini-implant anchorage. The complications of mini-implant anchorage mainly include structural damage during the implantation, peri-implantitis, mini-implant loosening, detachment, soft tissue damage, and root damage. Among them, peri-implantitis is the main complication leading to implant failure. The prevalence of peri-implantitis, an inflammatory condition affecting the tissues surrounding implants, leads to considerable patient discomfort, loosening or loss of mini-implants, and ultimately treatment failure. Peri-implantitis accounts for about 30% of mini-implant failures [1]. This highlights the necessity for effective strategies to enhance treatment outcomes in orthodontic practice.

Photobiomodulation therapy (PBMT) has emerged as a promising non-invasive approach to stimulate cellular metabolism and modulate biological processes without causing thermal damage. It refers to therapy involving the use of optical radiation typically in the visible and near-infrared spectrum to elicit photophysical and photochemical reactions for therapeutic benefits [2]. PBMT mainly includes low-level laser therapy (LLLT), light-emitting diode (LED) therapy, and broadband light therapy. Current studies have demonstrated that PBMT have potential biological stimulation effects on tissues, including promoting wound healing [3,4], relieving pain [5,6], promoting tissue regeneration and repair [7-9], and mitigating inflammation [9,10].

Current literatures also demonstrate a growing recognition of the great potential application of PBMT in the field of dentistry in terms of accelerating orthodontic tooth movement [11-15], mitigating associated pain [15,16], alleviates periodontal inflammation [17,18], inhibiting peri-implant inflammation and enhancing stability [19,20]. This non-invasive treatment offers a safe, efficient, and clinically proven approach in various dental procedures to improve overall patient care and recovery [21].

Previous studies indicated improved outcomes of PBMT in orthodontic applications, suggesting that LLLT or LED could effectively reduce mini-implants complications and enhance mini-implant stability [22]. However, there remains few literature reports regarding the specific effects of light-emitting diode laser application on the health of mini-implant and improve the success rate of mini-implant anchorage in orthodontic patients.

Peri-implant mucositis is a risk factor for mini-implantitis and mini-implant loosening and detachment [23]. The management of peri-implant mucositis is considered as a preventive measure for the onset of peri-implantitis to reduce the risk of treatment failure. This study employs a clinical randomized controlled trial to evaluate the effects of LED therapy on the management of peri-implant mucosa and mini-implant stability. It aims to provide a novel perspective on enhancing the stability and health of orthodontic mini-implants, ultimately improving treatment protocols and optimizing patient care.

## Materials and Methods

A total of thirty fixed orthodontic patients, aged 18-32 years, were selected, including 12 males and 18 females, who had their bilateral maxillary first premolars extracted for anterior teeth retraction. After alignment and leveling of the dentition, the arch wire was changed to 0.018x0.025inch stainless steel wire, and the self-tapping mini-implants (1.4mm*8mm, Vector TAs, Ormco Co.) were implanted at the alveolar membranogingival syndesmosis between the second premolars and the first molars on both sides of the maxilla as anchorage units. The left and right mini-implants of each patient were randomly assigned to the experimental group or the control group, with 30 cases in each group. Routine oral health education was carried out after the operation. The maxillary anterior teeth were retracted with a nickel-titanium spring of 150 grams between the mini-implants and the arch wire traction hook four weeks later (Fig. 1).

In the experimental group, a diode laser (wavelength 650nm, power 25mW) was used to irradiate the mucosal tissue around the neck of the mini-implant on days 0, 3, 7, and 14 after implantation, as well as on days 0, 3, 7, and 14 after mini-implant loading, respectively. The triangular area of the neck of the mini-implant was divided into three sites, with each site receiving irradiation for 20 seconds for a total of 1 minute (energy density 15.92 $J/cm^2$). The control group received simulated irradiation (no power irradiation). The plaque index, modified sulcus bleeding index, and probing depth of the mini-implants in both the experimental and control groups were evaluated at 1 week, 4 weeks, and 12 weeks after surgery. Cases of peri-implant mucositis and mini-implant loosening were recorded. The level of interleukin-1 β (IL-1 β) in the peri-implant fluid was detected by enzyme-linked immunosorbent assay (ELISA). The obtained data were analyzed using SPSS17.0 medical statistical software, and the counting data were statistically tested using the Fisher exact probability method. The measurement data results are expressed as the mean ± standard deviation ($\bar{x}$ ± s), and the paired t-test was used for the comparison of the means, with the test level of α = 0.05.

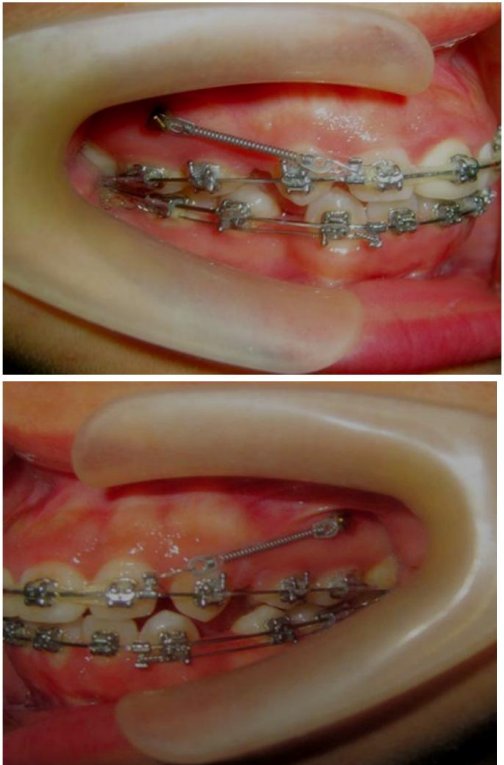

Figure 1　Implantation status of mini-implants

## Results

1. The mean values of the three detection indexes (plaque index, modified sulcus bleeding index, and probing depth) in the experimental group were lower than those in the control group at different time points, and the differences were statistically significant ($p<0.05$) (Fig. 2,3,4).
2. There was 1 case of peri-implant mucositis in the experimental group and 5 cases in the control group; the difference was statistically significant ($p<0.05$). Additionally, there were 2 cases of mini-implant loosening in the experimental group and 3 cases in the control group, with no statistically significant difference ($p>0.05$).
3. For IL-1 β, the concentration of IL-1 β in both groups was relatively high during the first week and then decreased. However, the concentration of IL-1 β in the experimental group was lower than that in the control group at each time point, with statistically significant differences ($p<0.05$) (Fig. 5).

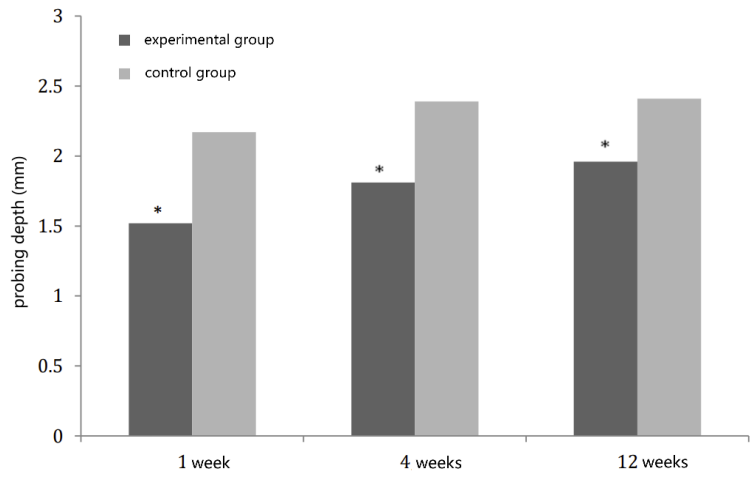

Figure 2 Comparison of probing depth between two groups at different time points.
* P<0.05

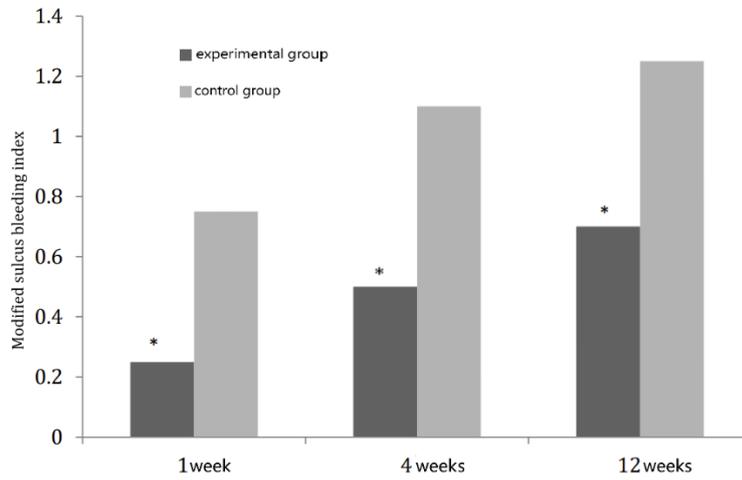

Fig. 3 Comparison of modified sulcus bleeding index between two groups at different time points.
* p<0.05

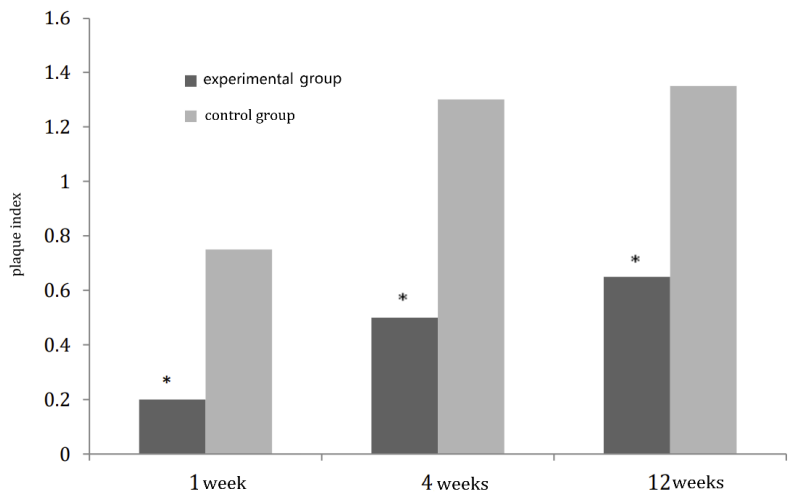

Fig. 4 Comparison of plaque index between two groups at different time points
* p<0.05

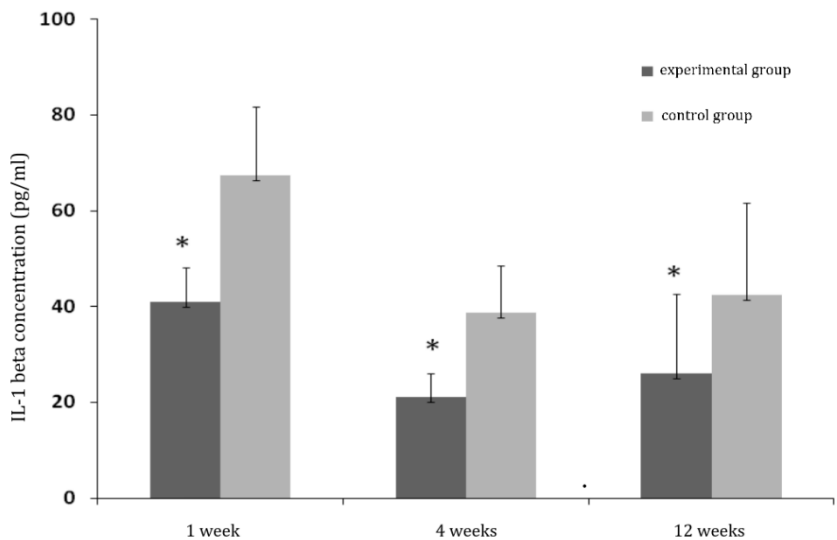

Figure 5    Comparison of IL-1β Concentrations Across Groups at Different Time Points
* p<0.05

## Discussion

At present, mini-implant anchorage has been widely used in various clinical treatments of orthodontics, playing an important role. The success of mini-implant in providing definitive anchorage depends on its stability and success rate in orthodontics. It is reported that the failure rates of orthodontic mini-implants varied between 3 and 19 percent after loading [24]. Peri-implantitis, characterized by inflammation of the tissues surrounding mini-implants, was turned out to be the main cause of mini-implant loosening [25,26]. The incidence of peri-implantitis can be attributed to a multitude of factors, including microbial infections, mechanical loading, host immune response and patient oral hygiene condition etc. Peri-implantitis is a progressive disease often triggered by bacterial invasion. Peri-implant mucositis, which is reversible inflammation of the mini-implant surrounding gingiva, is considered a precursor to peri-implantitis. It may progress to peri-implantitis if not treated. The management of peri-implant mucositis is considered as a preventive measure for the onset of peri-implantitis [23,27]. Understanding the pathogenesis of peri-implantitis is vital, as it lays the groundwork for developing effective management strategies that can enhance patient care and treatment success.

Mini-implant insertion causes an inflammatory reaction on gingival and bone tissues, which, in turn, triggers the biological processes associated with bone remodeling. Current studies present significant advancements in the management of orthodontic treatment complications associated with mini-implants, particularly through the application of diode laser therapy [22,28]. Photobiomodulation therapy may be useful in reducing inflammatory reactions and modulating bone remodeling, decreasing patient discomfort and the risk of mini-implant failure. This study aims to investigate the efficacy of LED therapy in mitigating complications associated with orthodontic mini-implants, particularly focusing on its impact on peri-implant tissue health and inflammation. Preliminary findings suggest that laser application significantly improves periodontal indices (plaque index, improved sulcus bleeding index, and probing depth) in the experimental group, reduces inflammation, and enhances overall implant stability compared to the control group. The outcomes are consistent with those reported in the literature [28,29]. These results underscore the potential of integrating laser therapy into orthodontic practices to optimize patient outcomes and minimize complications related to mini-implants.

This study focused on the complications associated with orthodontic mini-implants, specifically focusing on gingival conditions and changes in the concentration of lL-1 β in the surrounding fluid between the experimental group and the control group. Literature shows that in periodontal and peri-mini-implants tissues, cytokines may have essential roles in modulating inflammatory response. Several biochemical markers have been associated with inflammation and tissue remodeling, especially on osseo-integrated dental implants, but only few studies were found in the literature on the biochemical assessments of orthodontic mini-implants [30,31]. IL-1 β is a key pro-inflammatory cytokine that is often associated with gingivitis during orthodontic treatment [32]. Sehierano et al. [33] confirmed from molecular perspective that IL-1 β levels are associated with peri-implantitis. In this study, the concentration of IL-1 β in both groups were significantly higher during the first week and then decreased, which imply an early inflammatory reaction of the surrounding tissue to mini-implant insertion. The levels of IL-1 β gradually decreased at the 4th and 12th weeks. Moreover, the IL-1 β levels in the experimental group were statistically significantly lower than those in the control group at all timepoints, indicating the biological regulatory effect of PBMT. Greben [34] pointed out that PBMT regulate inflammatory response, reduce pro-inflammatory factors and promote tissue repair by affecting cell signaling pathways. The results of this study are consistent with the findings of Yassaei S [35]. Our research demonstrates the efficacy of diode laser photobiomodulation treatment in reducing inflammation and improving peri-implant tissue health in a human cohort.

In this study, the experimental group had fewer cases of peri-implant loosening than the control group but the difference was not statistically significant ($p > 0.05$). One possible reason is that our sample size was relatively small, and additionally, various factors can affect mini-implant primary stability, such as root proximity, surgery technique, loading time, mini-implant mucositis, and bone quality of insertion site [36]. It highlights the need for larger, further research with extended follow-up periods to validate and further elucidate the benefits of laser therapy in orthodontic practices.

The implications of our findings for clinical practice are considerable. The reduced incidence of peri-implant mucositis observed in patients receiving diode laser therapy underscores its potential as a standard adjunct in orthodontic treatments. This innovation not only enhances peri-implant health but may also lead to decreased treatment failures and improved patient comfort. Furthermore, these results align with current trends advocating for minimally invasive techniques in orthodontics.

The limitations of this study should be acknowledged including the relatively small sample size and the lack of long-term follow-up data. Additionally, the bio-stimulatory effect of LLLT is dose-dependent and different parameters of PBMT used in selected studies could also affect the reliability of the results. Further research is warranted to explore the biological effectiveness and refine laser-assisted orthodontic therapies.

In conclusion, this study demonstrates that adjunctive diode laser therapy can significantly reduce inflammatory markers and mitigate complications associated with mini-implants in orthodontic patients. The findings indicate the potential of PBMT as a supporting approach in orthodontic treatment to improve outcomes. Further research is warranted to explore the mechanisms of photobiomodulation and the best dosimetric radiation parameters for the successful application in clinical practice.